# Polarization dependence of photocurrent in a metal-graphene-metal device

Minjung Kim,[1] Ho Ang Yoon,[2] Seungwoo Woo,[1] Duhee Yoon,[1] Sang Wook Lee,[2] and Hyeonsik Cheong[1,a)]

[1]*Department of Physics, Sogang University, Seoul 121-742, Korea*
[2]*Division of Quantum Phases and Devices, School of Physics, Konkuk University, Seoul 143-701, Korea*

The dependence of the photocurrent generated in a Pd/graphene/Ti junction device on the incident photon polarization is studied. Spatially resolved photocurrent images were obtained as the incident photon polarization is varied. The photocurrent is maximum when the polarization direction is perpendicular to the graphene channel direction and minimum when the two directions are parallel. This polarization dependence can be explained as being due to the anisotropic electron-photon interaction of Dirac electrons in graphene.

Electronic and optoelectronic devices based on the unique physical properties of graphene are attracting much attention recently.[1–4] For optoelectronic applications, high-speed, broad-band photodetectors based on metal-graphene-metal junctions have been demonstrated.[5–11] Because graphene has a linear electronic band dispersion with no bandgap, light absorption is uniform over a wide range of the spectrum.[12–14] The high mobility of carriers and the large mean free path also allow high-speed operation of graphene photodetectors.[15,16] Although optical absorption in graphene, a hexagonal array of carbon atoms, is isotropic, the generation of photocurrent is expected to be polarization dependent due to the anisotropic nature of electron-photon interaction of the Dirac electrons in graphene.[17] So far, however, the polarization dependence of the graphene photodetectors has not been reported.

Grüneis $et\ al$. theoretically suggested that the optical absorption or emission probability of graphene is proportional to $\left| \vec{P} \times \vec{k} \right|^2$, where $\vec{P}$ is the polarization vector of a linearly polarized photon and $\vec{k}$ is the momentum vector of the electron excited by absorbing the photon.[17] This suggestion was invoked to explain the polarization dependence of the Raman $2D$ band observed in strained[18] and unstrained[19] single-layer graphene. However, a direct test of the theory has not been reported yet. In a metal-graphene-metal photodevice, the photocarriers excited by linearly polarized light, according to Ref. 17, should have momenta predominantly in the direction perpendicular to the polarization of the light. If the graphene channel is not long in comparison with the mean free path, partially ballistic transport of the photocarriers should lead to a polarization-dependent photocurrent. In this letter, we report the observation of the polarization dependence of the photocurrent generated in a Pd/graphene/Ti junction photodevice. Spatially resolved photocurrent images were measured as a function of the back-gate voltage and the polarization angle of the incident light. This observation is presented as direct evidence for the anisotropic electron-photon interaction suggested by Grüneis $et\ al$.

The graphene photodevice was fabricated by using the e-beam lithography process. Single-layer graphene samples were prepared on highly $p$-doped Si substrates covered with a 300-nm $SiO_2$ layer by mechanical exfoliation from natural graphite flakes. We identified the graphene sample to be single-layer using the shape of the Raman $2D$ band which has a single Lorentzian distribution.[20–22] As electrodes, a Pd (35 nm) electrode capped with Au (35 nm) and a Ti (40 nm) electrode capped with Au (40 nm) were deposited as shown in Fig. 1(a). Figure 1(b) shows a schematic of the spatially resolved photocurrent measurement setup used in this work. Raman and photocurrent images were obtained simultaneously in order to identify the exact position of the generated photocurrent in the graphene photodevice. Spatially resolved Raman and photocurrent images were obtained by raster-scanning a focused laser beam across the graphene photodevice. The 514.5-nm line of an Ar-ion laser, chopped at 100 Hz, was focused with a 50× objective lens (N.A. 0.8) to achieve a spatial resolution of approximately 700 nm, and the polarization was controlled by a half-wave plate. The laser power was kept at 150 μW, and all the measurements were carried out in ambient conditions. The photocurrent measurements were performed in the 2-probe configuration using the lock-in method for a better signal-to-noise ratio.

Figure 2(a) shows the photocurrent images as a function of the back-gate voltage $V_G$ from –50 V to 70 V. In the Pd-graphene junction, the generated photocurrent is negative for $V_G < -10$ V and positive for $V_G > -10$ V. In the graphene-Ti junction, on the other hand, the photocurrent is positive for $V_G < 0$ V and negative for $V_G > 0$ V. This behavior can be understood in terms of the band alignment shown in Figs. 2(b–d). Pd and Ti were chosen as the electrode materials in order to modify the work function of graphene in the region of the metal electrodes for maximum photocurrent.[9] At the charge neutrality point, the Dirac points of the graphene under the metal electrodes are higher than that in the channel region. The asymmetry between the Pd and Ti electrode is due to the difference in the doping of graphene induced by these metals.[23] In this case, photoexcited electrons near the electrodes are swept away from the electrodes owing to the build-in electric field due to band bending. This results in a positive current near the Pd electrode (source) and a negative current near the Ti electrode (grounded drain). As more negative gate bias is applied, the Dirac point of the channel region goes up in energy relative to the metal electrode regions. When the Dirac point of the channel region is between those in the electrode regions, as in Fig. 2(c), only positive currents are recorded throughout the channel region. When an even larger negative bias is applied, the Dirac point in the channel is higher than those in both of the electrode regions. In this case, the directions of the photocurrent become opposite to those at the charge neutrality point.

Figure 3(a) shows the polarization dependence of the photocurrent in the photodevice. We use the convention that the direction parallel to the edges of the electrode is 0 degree. The photocurrent images were measured as a function of the polarization angle of the incident laser. Because the measurements were performed in ambient conditions, the charge neutrality point drifts with time. In separate measurements, we found that the charge neutrality point moves by about 10 V during the first few minutes and then stabilizes. For the measurements, the back-gate voltage was initially set at +39 V and the photocurrent imaging measurements were taken after 20 minutes. After all the measurements were completed, the charge neutrality point was found at +49 V, as expected. Therefore, we can conclude that the measurements were performed in slightly p-doped conditions. The photocurrent is maximum when the polarization of the incident laser is 0, 180, and 360 degrees (Fig. 4). In order to ascertain that the observed polarization dependence is not due to some artifacts of the experimental set up, we rotated the sample by 90 degrees and repeated the measurements [Fig. 3(b)]. We keep the convention of setting the direction parallel to the edges of electrodes to 0 degree. Although there are some variations, the general trends are the same; the photocurrent is maximum when the polarization is parallel to the edges of the electrodes.

The polarization dependence results from the anisotropic electron-photon interaction in graphene.[17] Grüneis *et al*. calculated that the absorption of light by valence electrons in graphene is maximum when the polarization angle of the incident light is perpendicular to the momentum of the electron. If this prediction is correct, when the polarization of the incident light is parallel to the electrodes, absorption of light would be maximum for those electrons whose momenta are in the direction of the graphene channel. Since the momenta of electrons would be in the same direction as the internal electric field in this case, the photocurrent should be enhanced. On the other hand, if the polarization of the incident light is perpendicular to the electrodes, absorption of the light would be minimum for those electrons whose momenta are in the direction of the graphene channel, resulting in suppression of the photocurrent. It is evident that our result is consistent with this picture. We therefore conclude that the polarization dependence of the photocurrent in our device is direct evidence for the anisotropic electron-photon interaction suggested by Grüneis *et al.*[17]

However, the variation of the photocurrent $(I_{max} - I_{min})/I_{max} = 0.18 \sim 0.26$ is not very large, which means that the photocurrent is not perfectly polarized. If the generated photocurrent is ballistic, the polarization should be much larger. However, the scattering of carriers due to the charged centers in the $SiO_2$ substrate should reduce the portion of ballistic transport of the photocarriers, resulting in a small polarization ratio.[24–26] A higher polarization ratio might be possible if more inert substrates such as hexagonal boron nitride is used.

In summary, we measured the polarization dependence of the photocurrent in a Pd/graphene/Ti photodevice. The photocurrent is maximum when the polarization of incident light is perpendicular to the channel direction of the graphene photodevice, which is direct evidence for anisotropic electron-photon interaction of Dirac electrons in graphene. This effect might be utilized for the development of polarization sensitive photodetectors.

We thank Prof. W. Y. Lee of Yonsei University for making a wire bonder available for sample preparation. This work was supported by the National Research Foundation of Korea grant funded by the Ministry of Education, Science and Technology of Korea (No. 2011-0017605 and No. 2011-0021207), WCU (R31-2008-000-10057-0), and a grant (No. 2011-0031630) from the Center for Advanced Soft Electronics under the Global Frontier Research Program of the MEST.

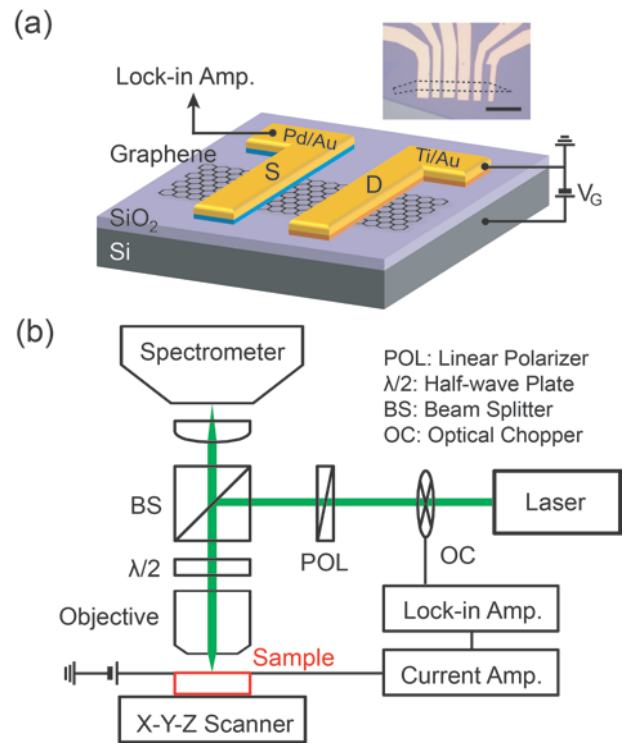

**FIG. 1.** (a) Schematic diagram of the graphene photodevice. The inset is an optical microscope image of the sample. The scale bar is 20 μm. The distance between the electrodes is 1.5 μm. (b) Schematic diagram of the spatially resolved photocurrent and Raman measurements setup.

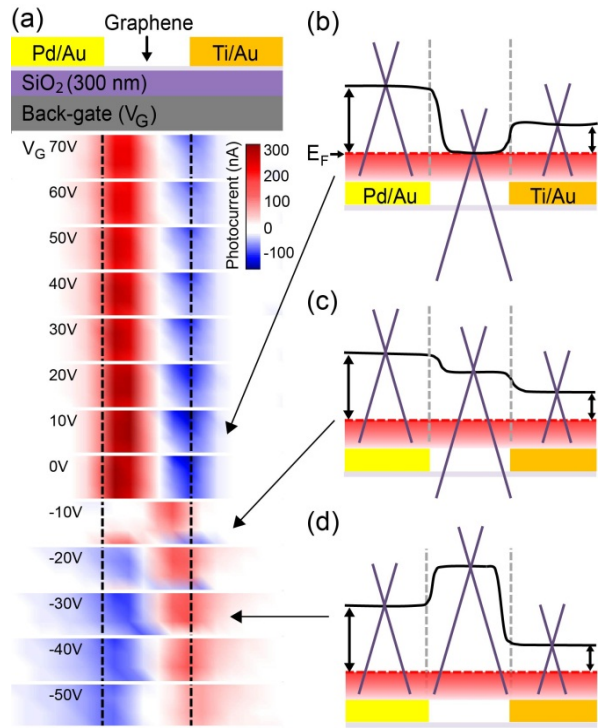

**FIG. 2.** (a) Photocurrent images as a function of the back-gate voltage $V_G$. (b–d) Surface potential profiles of the graphene photodevice for various values of $V_G$. $E_F$ is the Fermi energy.

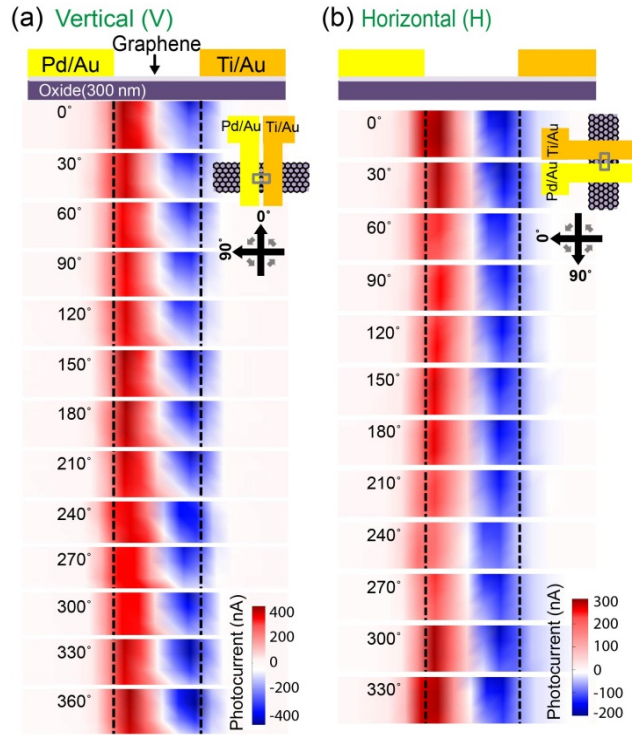

**FIG. 3.** Photocurrent images as a function of the polarization angle of the incident light for (a) vertical (V) and (b) horizontal (H) configurations. The direction parallel to the edges of the electrodes is defined 0˚ in both the configurations.

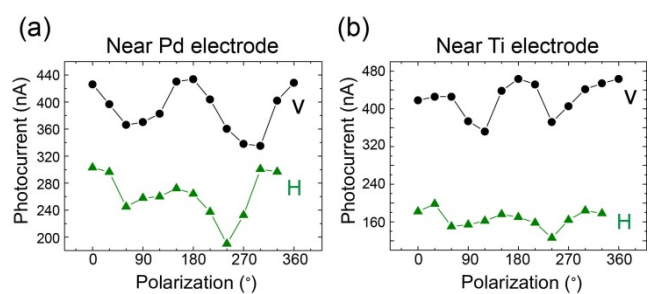

**FIG. 4.** Polarization dependence of the magnitude of the photocurrent near (a) the Pd and (b) the Ti electrode. The data for both V and H configurations are plotted.